\documentclass[%
 reprint,
superscriptaddress,%groupedaddress,%unsortedaddress,%runinaddress,%frontmatterverbose,%preprint,%showpacs,preprintnumbers,%nofootinbib,%nobibnotes,%bibnotes,
amsmath,amssymb,
aps,
pra,
]{revtex4-1}

\usepackage{graphicx}% Include figure files
\usepackage{dcolumn}% Align table columns on decimal point
\usepackage{bm}% bold math

\begin{document}

\preprint{APS/123-QED}

\title{Non-diffracting broadband incoherent space-time fields}

\author{Murat Yessenov}
\author{Basanta Bhaduri}
\author{H. Esat Kondakci}
\affiliation{CREOL, The College of Optics \& Photonics, University of Central Florida, Orlando, Florida 32186, USA}

\author{Monjurul Meem}
\author{Rajesh Menon}
\affiliation{Department of Electrical \& Computer Engineering, University of Utah, Salt Lake City, UT 84112, USA}

\author{Ayman F. Abouraddy}
 \email{Corresponding author: raddy@creol.ucf.edu}
 \affiliation{CREOL, The College of Optics \& Photonics, University of Central Florida, Orlando, Florida 32186, USA}

% To be edited by editor
% \dates{Compiled \today}

% To be edited by editor
% \doi{\url{http://dx.doi.org/10.1364/optica.XX.XXXXXX}}

\begin{abstract}
\normalsize{Space-time (ST) wave packets are coherent pulsed beams that propagate diffraction-free and dispersion-free by virtue of tight correlations introduced between their spatial and temporal spectral degrees of freedom. Less is known of the behavior of \textit{in}coherent ST fields that maintain the spatio-temporal spectral {structure} of their coherent wave-packet counterparts while losing all purely spatial or temporal coherence. We show here that structuring the spatio-temporal spectrum of an incoherent field produces broadband incoherent ST fields that are diffraction free. The intensity profile of these fields consist of a narrow spatial feature atop a constant background. Spatio-temporal spectral engineering allows controlling the width of this spatial feature, tuning it from a bright to a dark diffraction-free feature, and varying its amplitude relative to the background. These results pave the way to new opportunities in the experimental investigation of optical coherence of fields jointly structured in space and time by exploiting the techniques usually associated with ultrafast optics.}
\end{abstract}

\maketitle

%\begin{document}

\section{Introduction}

Diffraction-free beams have a long history \cite{Rayleigh72MNRAS,Sheppard77Optik} culminating in the decisive demonstration by Durnin \textit{et al.} of a monochromatic Bessel beam \cite{Durnin87PRL}, among others that share its extended diffraction length \cite{Levy16PO}. \textit{Pulsed} beams that are simultaneously diffraction-free \textit{and} dispersion-free have also been identified, including Brittingham's focus wave mode \cite{Brittingham83JAP}, MacKinnon's wave packet \cite{Mackinnon78FP}, and X-waves \cite{Lu92IEEEa,Saari97PRL}, among other possibilities \cite{Turunen10PO,FigueroaBook14}. These propagation-invariant pulsed beams have been recently investigated under the collective moniker of `space-time' (ST) wave packets because a fundamental feature underpins them all: the spatial and temporal frequencies underlying the beam profile and pulse linewidth, respectively, are tightly correlated \cite{Donnelly93PRSLA,Longhi04OE,Saari04PRE,Kondakci16OE,Parker16OE}. Spatial frequency refers to the transverse component of the wave vector and temporal frequency to the usual (angular) frequency, which also entails also defining spatial and temporal bandwidths. We adopt this unfamiliar nomenclature to symmetrize the treatment of the spatial and temporal spectral degrees of freedom. Utilizing this principle, we have devised an efficient synthesis strategy for implementing arbitrary one-to-one relationships between the spatial and temporal frequencies \cite{Kondakci17NP}. This has led to demonstrations of non-accelerating Airy wave packets \cite{Kondakci18PRL}, self-healing \cite{Kondakci18OL}, extended propagation distances \cite{Bhaduri18OE,Bhaduri19OL} and group delays \cite{Yessenov19OL}, arbitrary group velocities in free space \cite{Kondakci18unpub} and non-dispersive materials \cite{Bhaduri19Optica}, and tilted-pulse-front ST wave packets \cite{Kondakci18ACSP}.

Less attention has been devoted to \textit{incoherent} propagation-invariant fields, especially broadband fields that are incoherent in space \textit{and} time \cite{Turenen91JOSAA,Friberg91PRA,Bouchal92JMO,Fischer05OE,Fischer06JOA,Turunen08OE,Saastamoinen09PRA}. Here we present, to the best of our knowledge, the first experimental demonstration of broadband incoherent ST fields in which precise control is exercised over the spatio-temporal spectral content, leading to several new observations. Indeed, our results oppose the usual expectation that diffraction-free behavior is diminished once spatial and temporal coherence are both reduced. To best frame our results in context, we first provide a brief summary of previous achievements, followed by the specific contributions of this paper.

Initial work on propagation-invariant incoherent fields investigated the impact of \textit{spatial} coherence on the propagation of quasi-monochromatic fields. In Ref.~\cite{Turenen91JOSAA}, starting with a monochromatic spatially coherent laser transformed into a Bessel beam via a holographic plate, a scattering diffuser was placed in the beam focal plane, whereupon an axially invariant speckle pattern and coherence function were observed \cite{Turenen91JOSAA}. A subsequent theoretical study introduced quasi-monochromatic spatially incoherent `dark' and `anti-dark' diffraction-free beams consisting of a uniform-intensity incoherent background superposed with a spatially narrow peak or dip \cite{Ponomarenko07OL}, which have not been observed to date. Along similar lines, spatially incoherent Airy beams have been examined. After an initial report suggested that spatial coherence reduces the range over which acceleration is observed \cite{Morris09OE}, a subsequent study showed that the diminishing of self-acceleration induced by incoherence can be prevented by ensuring that all the constituent coherent modes accelerate along the same trajectory \cite{Lumer15Optica}. That study utilized a monochromatic laser, and spatial incoherence was introduced by temporally modulating a phase mask or via a rotating diffuser. 

The impact of spatial \textit{and} temporal incoherence on the propagation of a Bessel beam produced by a variety of sources was examined experimentally in Ref.~\cite{Fischer05OE}, with the conclusion that spatial coherence reduces the diffraction-free length. However, joint control over the spatial and temporal degrees of freedom was \textit{not} exercised in that experiment. Finally, theoretical studies exploiting new representations of incoherent spatio-temporal fields have examined the impact of introducing \textit{partial} coherence into propagation-invariant wave packets when the spatial and temporal frequencies are correlated  \cite{Turunen08OE,Saastamoinen09PRA}. In summary, there have been no observations of broadband incoherent fields endowed with the spatio-temporal spectral correlations underpinning their coherent pulsed counterparts.

Here we exploit a light-emitting diode (LED) to synthesize diffraction-free broadband incoherent fields in which the spatial and temporal frequencies are tightly correlated but the spectral amplitudes derived from the LED lack mutual coherence. By combining techniques from ultrafast pulse shaping with spatial beam modulation applied to incoherent light, we efficiently synthesize new \textit{incoherent ST fields}. Unprecedented precision is achieved in the synthesis process, where the 20-nm bandwidth of the LED is modulated in conjunction with the spatial degree of freedom with a precision of $\sim25$~pm. We confirm that the spatial intensity profile is indeed invariant for more than two-orders of magnitude ($\sim\!300\times$) of the Rayleigh range. These fields enable the exploration of a new regime of optical coherence where spatial and temporal coherence are both lacking -- but the spatial and temporal spectral degrees of freedom are inextricably intertwined. The spatial intensity profile of an \textit{in}coherent ST field is \textit{indistinguishable} from its coherent wave-packet counterpart, and comprises a sharply defined peak atop of a broad background. By manipulating the spectral \textit{amplitudes}, we control the width of the narrow peak and its ratio to the constant background; and by modifying the spectral \textit{phases}, we convert the narrow peak to a dip. For simplicity we deal with one-dimensional (1D) fields in the form of light sheets while holding the intensity uniform along the other transverse dimension. Pioneering efforts by Saari in the synthesis of X-waves \cite{Saari97PRL} and focus-wave modes \cite{Reivelt02PRE} also made use of incoherent light. The fields we synthesize here are so-called `baseband' ST fields in which the low spatial frequencies -- which are excluded from X-waves and focus-wave modes -- are retained, thus making the synthesis of ST fields and control over their properties considerably easier \cite{Yessenov18PRA}.

We emphasize that the concept of spatio-temporal `spectral correlations' refers to the tight \textit{deterministic} association between spatial and temporal frequencies according to a prescribed functional form -- and \textit{not} to any form of \textit{statistical} correlations. Each spatial frequency is associated with a single wavelength within a very narrow window of spectral uncertainty ($\sim\!25$~pm) \cite{Kondakci19OL,Yessenov19OL}.

The paper is organized as follows. First, we present a brief description of coherent ST wave packets highlighting the role of their tight spatio-temporal spectral correlations in enforcing propagation invariance. Next, we describe the conditions for propagation invariance of a spatially incoherent quasi-monochromatic field, before extending the theoretical analysis to broadband fields. We then present the experimental arrangement for producing diffraction-free incoherent ST fields from a LED, and then present our experimental findings, including the effect of varying the amplitude and phase of the complex spatio-temporal spectral coherence function. Finally, we discuss the implications and future extensions of this work before presenting our conclusions. 

\begin{figure*}[t!]
\centering
\includegraphics[width=17.2cm]{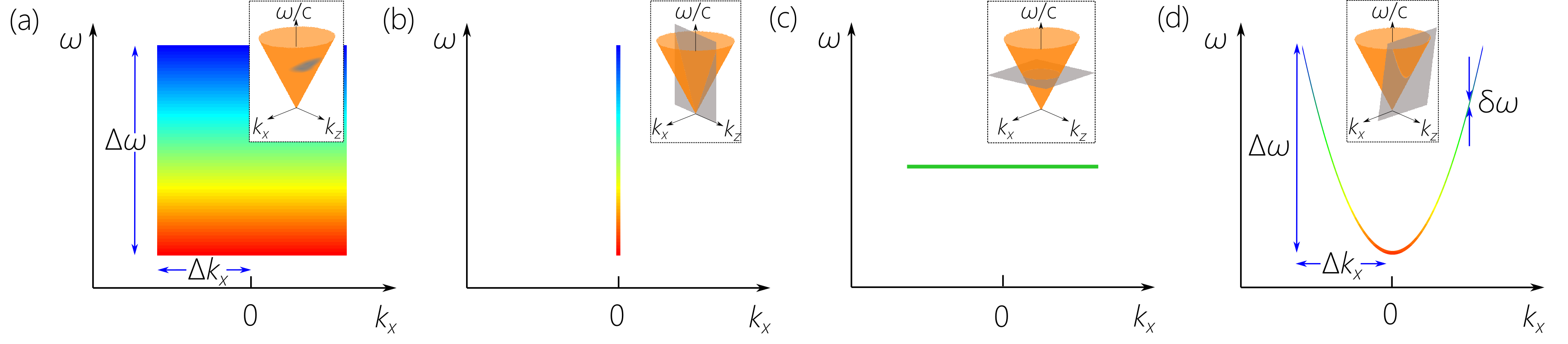}
\caption{\small{Concept of diffraction-free coherent and incoherent space-time (ST) fields. Each panel depicts a spatio-temporal spectrum in the $(k_{x},\omega)$-plane. For \textit{coherent} fields, the spectra should be interpreted as spectral densities $|\widetilde{E}(k_{x},\omega)|^{2}$. For \textit{in}coherent fields, the spectral coherence function $\widetilde{G}(k_{x},\omega;k_{x}',\omega')$ in Eq.~\ref{Eq:GeneralCoherenceFunction} is cast as $\widetilde{G}(k_{x},\omega)\delta(\omega-\omega')\delta(|k_{x}|-|k_{x}'|)$, and the spectra plotted here correspond to $\widetilde{G}(k_{x},\omega)$. Insets show the spatio-temporal spectral configurations on the surface of the light-cone $k_{x}^{2}+k_{z}^{2}\!=\!(\tfrac{\omega}{c})^{2}$. (a) A typical spectrum separable with respect to the spatial and temporal degrees of freedom; $\Delta k_{x}$ and $\Delta\omega$ are the spatial and temporal bandwidths, respectively. (b) Spatial filtering of the spectrum in (a) yields a plane wave with finite temporal bandwidth; corresponding to the intersection of the light-cone with the plane $k_{x}\!=\!0$. (c) Temporal filtering of the spectrum in (a) yields a quasi-monochromatic field with finite spatial bandwidth; corresponding to the intersection of the light-cone with a plane $\omega\!=\!\omega_{\mathrm{o}}$. (d) Spatio-temporal filtering of the spectrum in (a) yields a curved spectral trajectory that we refer to as an ST field; corresponding to the intersection of the light-cone with a tilted plane parallel to the $k_{x}$-axis. If this trajectory satisfies the constraint in Eq.~\ref{Eq:CorrelationFunction} with coherent light, then we have a propagation-invariant pulsed beam or wave packet; otherwise we have a diffraction-free broadband incoherent ST field. Here $\delta\omega$ is the spectral uncertainty in the association between the spatial and temporal frequencies.}}
\label{Fig:Concept}
\end{figure*}

\section{Coherent space-time wave packets}

For coherent 1D fields in the form of pulsed light sheets $E(x,z,t)$, the spatio-temporal spectrum is spanned by the spatial frequency $k_{x}$ and temporal frequency $\omega$, where $x$ is a transverse coordinate, and we assume the field is uniform along the other transverse coordinate $y$. In free space, the dispersion relation $k_{x}^{2}+k_{z}^{2}\!=\!(\tfrac{\omega}{c})^{2}$ corresponds geometrically to the surface of the light-cone, where $k_{z}$ is the axial component of the wave vector along the axial coordinate $z$, and $c$ is the speed of light in vacuum (see Fig.~\ref{Fig:Concept} insets). The spatio-temporal spectrum of a typical pulsed beam (or wave packet) has finite spatial \textit{and} temporal bandwidths $\Delta k_{x}$ and $\Delta\omega$, respectively, and is thus represented by a 2D patch on the surface of the light-cone. Figure~\ref{Fig:Concept}(a) illustrates schematically such a spectrum, which is usually assumed to be separable in $k_{x}$ and $\omega$. Such a beam diffracts upon propagation, and space-time coupling renders the wave packet subsequently non-separable in space and time \cite{SalehBook07}. Filtering such a wave packet spatially, while retaining the temporal bandwidth $\Delta\omega$, produces a pulsed plane wave [Fig.~\ref{Fig:Concept}(b)]; whereas filtering the temporal spectrum while retaining its spatial bandwidth $\Delta k_{x}$ results in a monochromatic beam [Fig.~\ref{Fig:Concept}(c)].

The spatio-temporal spectrum of a propagation-invariant wave packet (diffraction-free \textit{and} dispersion-free) takes the form of a 1D curved trajectory as depicted in Fig.~\ref{Fig:Concept}(d), which can be produced via one of three pathways. First, by spatio-temporal filtering of the generic spectrum in Fig.~\ref{Fig:Concept}(a) and introducing one-to-one relationship between $\omega$ and $|k_{x}|$ with a spectral uncertainty of $\delta\omega$ -- while retaining the bandwidths $\Delta k_{x}$ and $\Delta\omega$. Second, the configuration in Fig.~\ref{Fig:Concept}(d) can be reached from the pulsed plane wave in Fig.~\ref{Fig:Concept}(b) by wavelength-dependent spatial beam modulation (a spatio-temporal Fourier-optics approach) to increase the spatial bandwidth to $\Delta k_{x}$. A third strategy starts from the monochromatic beam in Fig.~\ref{Fig:Concept}(c) and utilizes nonlinear optics to introduce a $k_{x}$-dependent frequency shift and increase the temporal bandwidth to $\Delta\omega$. We make use of the second approach that is particularly apt for incoherent light because it requires only linear optics.

Propagation invariance requires that the curved spatio-temporal spectral trajectory in Fig.~\ref{Fig:Concept}(d) be a conic section resulting from the intersection of the light-cone with a tilted spectral hyperplane $\mathcal{P}(\theta)$ described by the equation
\begin{equation}\label{Eq:CorrelationFunction}
\tfrac{\omega}{c}\!=\!k_{\mathrm{o}}+(k_{z}-k_{\mathrm{o}})\tan{\theta},
\end{equation}
where $k_{\mathrm{o}}\!=\!\omega_{\mathrm{o}}/c$ is a fixed wave number and $\theta$ is the spectral tilt angle of $\mathcal{P}$ with respect to the $(k_{x},k_{z})$-plane \cite{Yessenov18PRA}. After applying this constraint, the field expressed as a product of an envelope and optical carrier $E(x,z,t)\!=\!e^{i(k_{\mathrm{o}}z-\omega_{\mathrm{o}}t)}\psi(x,z,t)$ becomes
\begin{equation}\label{Eq:STWavePacket}
E(x,z,t)\!=\!e^{i(k_{\mathrm{o}}z-\omega_{\mathrm{o}}t)}\!\!\!\int\!\! dk_{x}\widetilde{\psi}(k_{x})e^{ik_{x}x}e^{-i(\omega-\omega_{\mathrm{o}})(t-\frac{z}{c}\cot{\theta})},
\end{equation}
and the envelope $\psi(x,z,t)\!=\!\psi(x,0,t-\frac{z}{c}\cot{\theta})$ is a pulsed beam transported rigidly along $z$ at a group velocity $v_{\mathrm{g}}\!=\!c\tan{\theta}$ \cite{Kondakci18unpub}; here $\widetilde{\psi}(k_{x})$ is the Fourier transform of $\psi(x,0,0)$. The frequency $\omega\!=\omega(|k_{x}|)$ in Eq.~\ref{Eq:STWavePacket} is tightly correlated with the spatial frequency $k_{x}$. A quadratic relationship exists between $\omega$ and $k_{x}$ corresponding to a conic section resulting from the intersection of the plane described in Eq.~\ref{Eq:CorrelationFunction} with the light-cone $k_{x}^{2}+k_{z}^{2}\!=\!(\tfrac{\omega}{c})^{2}$ \cite{Kondakci17NP,Kondakci18unpub,Yessenov18PRA}. We refer to such pulsed beams as ST wave packets, which we have recently synthesized in Refs.~\cite{Kondakci17NP,Kondakci18PRL,Kondakci18OE,Kondakci18OL,Bhaduri18OE,Kondakci18unpub,Kondakci18ACSP}. The spatio-temporal intensity profile of this wave packet is $I(x,z,t)\!=\!|E(x,z,t)|^{2}\!=\!|\psi(x,0,t-\tfrac{z}{c}\cot{\theta})|^{2}$, and the time-averaged intensity $I(x,z)\!=\!\int\!dtI(x,z,t)$ as observed with a `slow' detector, such as a CCD camera, is given by
\begin{equation}
I(x,z)\!=\!I(x,0)\!=\!\!\!\int\!\!\!dk_{x}|\widetilde{\psi}(k_{x})|^{2}\!+\!\!\int\!\!\!dk_{x}\;\widetilde{\psi}(k_{x})\widetilde{\psi}^{*}(-k_{x})e^{i2k_{x}x}\!,
\end{equation}
which simplifies to $I(x,z)\!=\!\int\!\!dk_{x}|\widetilde{\psi}(k_{x})|^{2}(1\!+\!\cos{2k_{x}x})$ when $\widetilde{\psi}$ is an even function. That is, the time-averaged intensity takes the form of a narrow spatial feature atop a constant background. To the best of our knowledge, the impact of introducing the spatio-temporal correlations governed by Eq.~\ref{Eq:CorrelationFunction} into a broadband \textit{incoherent} field has \textit{not} been investigated experimentally to date.

\section{Diffraction-free quasi-monochromatic partially coherent fields}

Under what conditions does a 1D \textit{quasi-monochromatic} field that is spatially partially coherent become diffraction-free? The coherence function $G(x_{1},z;x_{2},z)\!=\!\langle E(x_{1},z)E^{*}(x_{2},z)\rangle$ at a plane $z$ is represented in the spatial-frequency domain as follows:
\begin{equation}
G(x_{1},z;x_{2},z)\!\!=\!\!\!\int\!\!\!\!\int\!\!\!dk_{x}dk_{x}'\;\!\widetilde{G}(k_{x},k_{x}')e^{i(k_{x}x_{1}-k_{x}'x_{2})}e^{i(k_{z}-k_{z}')z}\!,
\end{equation}
where $\widetilde{G}(k_{x},k_{x}')$ is a spatial-frequency correlation function and the Fourier transform of $G(x_{1},0;x_{2},0)$. Enforcing propagation invariance $G(x_{1},z;x_{2},z)\!=\!G(x_{1},0;x_{2},0)$ can be shown to imply the constraint $\widetilde{G}(k_{x},k_{x}')\rightarrow\widetilde{G}(k_{x},k_{x}')\delta(|k_{x}|-|k_{x}'|)$ such that
\begin{equation}\label{Eq:G1G2SpectralDecomposition}
\widetilde{G}(k_{x},k_{x}')=\widetilde{G}_{1}(k_{x})\delta(k_{x}-k_{x}')+\widetilde{G}_{2}(k_{x})\delta(k_{x}+k_{x}'),
\end{equation}
where $\widetilde{G}_{1}(k_{x})\!=\!\widetilde{G}(k_{x},k_{x})$ and $\widetilde{G}_{2}(k_{x})\!=\!\widetilde{G}(k_{x},-k_{x})$. Consequently, all the spatial frequencies must be mutually incoherent as seen from the $\widetilde{G}_{1}(k_{x})$-term, with the potential exception of the $\pm k_{x}$ pairs, which can be correlated if the $\widetilde{G}_{2}(k_{x})$-term is non-zero. Note that $\widetilde{G}_{1}$ is real and positive, whereas $\widetilde{G}_{2}$ may be complex, $\widetilde{G}_{2}(k_{x})\!=\!|\widetilde{G}_{2}(k_{x})|e^{i\varphi(k_{x})}$ with $\varphi(-k_{x})\!=\!-\varphi(k_{x})$. Therefore, propagation invariance of the coherence function enforces a reduction of its dimensionality from 2D to 1D. The propagation-invariant coherence function can now be written as a sum,
\begin{equation}
G(x_{1},x_{2})=G_{1}(x_{1}-x_{2})+G_{2}(x_{1}+x_{2}),
\end{equation}
where $G_{1}(x)\!=\!\int\!dk_{x}\widetilde{G}_{1}(k_{x})e^{ik_{x}x}$, $G_{2}(x)\!=\!\int\!dk_{x}\widetilde{G}_{2}(k_{x})e^{ik_{x}x}$, $|G_{2}(x)|\!\leq\!|G_{1}(0)|$ for all $x$, and the intensity $I(x)\!=\!G(x,x)$ is
\begin{equation}\label{Eq:IntensityQuasiMonochromatic}
\begin{split}
I(x)\!=\!G_{1}(0)\!+\!G_{2}(2x)\!= \\
\!G_{1}(0)\!+ \!\!\int\!\!dk_{x}|\widetilde{G}_{2}(k_{x})|\cos{(2k_{x}x+\varphi(k_{x}))}.
\end{split}
\end{equation}

It is critical to appreciate the physical significance of the two terms $G_{1}$ and $G_{2}$. If \textit{all} the spatial frequencies are mutually incoherent, then $G_{2}(x)\!=\!0$, resulting in a uniform intensity profile $I(x)\!=\!G_{1}(0)$, which corresponds to the experiment in Ref.~\cite{Turenen91JOSAA}. If the field amplitudes associated with the spatial frequencies $k_{x}$ and $-k_{x}$ are equal, fully correlated, and $\varphi(k_{x})\!=\!0$ while remaining mutually incoherent with all other spatial frequencies, then the propagation-invariant intensity takes the form of a constant background $G_{1}(0)$ superposed with an \textit{equal-amplitude} shaped beam $G_{2}(2x)$ that can be extremely narrow. However, if the field amplitudes at $k_{x}$ and $-k_{x}$ are anti-correlated such that $\varphi(k_{x})$ is a $\pi$-step along $k_{x}$, and $G_{2}$ is therefore negative, then $I(x)$ features a non-diffracting \textit{dip} in the intensity profile. This configuration corresponds to the `dark' and `anti-dark' fields studied theoretically in Ref.~\cite{Ponomarenko07OL} (and also hinted at in Ref.~\cite{Turenen91JOSAA}), which have \textit{not} been realized to date.

\section{Theory of spatio-temporal coherence}

We now proceed to stochastic fields in space \textit{and} time whose coherence function $G(x_{1},z_{1},t_{1};x_{2},z_{2},t_{2})\!=\!\langle E(x_{1},z_{1},t_{1})E^{*}(x_{2},z_{2},t_{2})\rangle$ expressed in the spectral domain is
\begin{equation}\label{Eq:GeneralCoherenceFunction}
\begin{split}
G(x_{1},z_{1},t_{1};x_{2},z_{2},t_{2}\!)\!\!=\!\!\!\int\!\!\!\!\int\!\!dk_{x}dk_{x}'\!\!\int\!\!\!\!\int\!\!d\omega d\omega'\,\,\!\!\widetilde{G}(k_{x},\omega;k_{x}',\omega')\\ e^{i(k_{x}x_{1}-k_{x}'x_{2})}e^{i(k_{z}z_{1}-k_{z}'z_{2})}e^{-i(\omega t_{1}-\omega't_{2})}\!,
\end{split}
\end{equation}

where the triplets $(k_{x},k_{z},\tfrac{\omega}{c})$ and $(k_{x}',k_{z}',\tfrac{\omega'}{c})$ independently satisfy the free-space dispersion relationship. Our theoretical analysis here is formulated in the spatio-temporal spectral domain $(k_{x},\omega)$ -- as opposed to physical space $(x,z,t)$ or the space-frequency domain $(x,z,\omega)$ \cite{Saastamoinen09PRA} -- because it brings out clearly the spatio-temporal spectral correlation structure and thus the reduced-dimensionality intrinsic to its coherent counterparts. Furthermore, as described below, examining the field in the $(k_{x},\tfrac{\omega}{c})$-domain also immediately suggests a methodology for synthesizing incoherent ST fields through the manipulation of their spatio-temporal spectrum. The spatio-temporal spectral coherence function $\widetilde{G}(k_{x},\omega;k_{x}',\omega')$ is four-dimensional (six-dimensional if $y$ is retained), which presents a daunting prospect for analysis. We consider here two relevant limits that help simplify the analysis: a spectrum that is separable in space and time, and one in which the tight spatio-temporal correlations of Eq.~\ref{Eq:CorrelationFunction} are introduced. 

First, in the special case of a field that has a \textit{separable} spatio-temporal spectrum $\widetilde{G}(k_{x},\omega;k_{x}',\omega')\!=\!\widetilde{G}_{x}(k_{x},k_{x}')\widetilde{G}_{t}(\omega,\omega')$, the coherence function $G$ in turn separates in the input plane $z_{1}\!=\!z_{2}\!=\!0$, $G(x_{1},0,t_{1};x_{2},0,t_{2})\!=\!G_{x}(x_{1},x_{2})G_{t}(t_{1},t_{2})$; that is, an optical field that is cross-spectrally pure \cite{Mandel61JOSA,Mandel76JOSA}. This scenario corresponds to the illustration in Fig.~\ref{Fig:Concept}(a), where each spatio-temporal spectral amplitude corresponds to a random variable that is uncorrelated with all the others. As the field propagates, space-time coupling renders the coherence function \textit{non}-separable for $z\!>\!0$, and cross-spectral purity is gradually lost. Spatial filtering produces a temporally incoherent plane wave [Fig.~\ref{Fig:Concept}(b)], whereas temporal filtering produces a quasi-monochromatic spatially incoherent field [Fig.~\ref{Fig:Concept}(c)].

Now consider the other limit where $k_{x}$ and $\omega$ are tightly correlated [Fig.~\ref{Fig:Concept}(d)] according to the relationship in Eq.~\ref{Eq:CorrelationFunction}. In other words, $\widetilde{G}(k_{x},\omega;k_{x}',\omega')\!\rightarrow\!\widetilde{G}(k_{x},k_{x}')\delta(\omega-f(|k_{x}|))\delta(\omega'-f(|k_{x}'|))$, where $\omega\!=\!f(|k_{x}|)$ is the locus of the spatio-temporal spectrum along the conic section at the intersection of the light-cone with the spectral hyperplane $\mathcal{P}(\theta)$. In the coherent case, this constraint produces a propagation-invariant ST wave packet as given in Eq.~\ref{Eq:STWavePacket}. In the incoherent case considered here, the coherence function takes the form
\begin{equation}\label{Eq:PropagationOfBroadbandCoherence}
\begin{split}
G(x_{1},z_{1},t_{1};x_{2},z_{2},t_{2})\!\!=\!\!\!\int\!\!\!\int\!dk_{x}dk_{x}'\;\widetilde{G}(k_{x},k_{x}')e^{i(k_{x}x_{1}-k_{x}'x_{2})} \\ e^{-i(\omega-\omega_{\mathrm{o}})(t_{1}-\frac{z_{1}}{c}\cot{\theta})}
e^{i(\omega'-\omega_{\mathrm{o}})(t_{2}-\frac{z_{2}}{c}\cot{\theta})},
\end{split}
\end{equation}
whereupon the dimensionality of the spatio-temporal spectral coherence function $\widetilde{G}$ has dropped from 4D to 2D because each spatial frequency $k_{x}$ is assigned to a single temporal frequency $\omega$. The coherence function in a single plane $z$ is then given by
\begin{equation}
\begin{split}
G(x_{1},z,t_{1};x_{2},z,t_{2})=e^{i\omega_{\mathrm{o}}(t_{1}-t_{2})} \int\!\!\!\int\!dk_{x}dk_{x}'\;\widetilde{G}(k_{x},k_{x}')\\ e^{i(k_{x}x_{1}-k_{x}'x_{2})}   e^{-i(\omega t_{1}-\omega't_{2})}e^{i\frac{z}{c}\cot{\theta}(\omega-\omega')}.
\end{split}
\end{equation}

The strategy employed in synthesizing the incoherent ST field governs the form of the spectral coherence function $\widetilde{G}(k_{x},k_{x}')$. In our approach, we make use of a broadband temporally incoherent field from a LED whose spatio-temporal spectrum corresponds to that in Fig.~\ref{Fig:Concept}(a), for which we can assume the usual delta-function correlation in $\omega$; i.e., all the temporal frequencies are mutually incoherent. We then spatially filter this field and reach the configuration in Fig.~\ref{Fig:Concept}(b), whereupon the field is spatially coherent but temporally incoherent. Subsequently we judiciously assign a pair of spatial frequencies $\pm k_{x}$ to each temporal frequency $\omega$, such that we reach the sought-after configuration in Fig.~\ref{Fig:Concept}(d). At this stage, the field possesses \textit{neither} temporal \textit{nor} spatial coherence. That is, despite starting from a spatially filtered field, once the spatial and temporal frequencies are correlated with each other $\omega\!=\!f(|k_{x}|)$, the temporal incoherence is shared and inherited by the spatial degree of freedom. The spectral amplitudes associated with different spatial frequencies are now mutually incoherent, except for $\pm k_{x}$ that are related because they are both derived from same field amplitude at $\omega$ in the spatially filtered field in Fig.~\ref{Fig:Concept}(b). As a result, all the spatial frequencies in turn have a delta-function correlation \textit{except} for the pairs $\pm k_{x}$, such that 
$\widetilde{G}(k_{x},k_{x}')\!\rightarrow\!\widetilde{G}(k_{x})\delta(|k_{x}|-|k_{x}'|)$ with the concomitant decomposition in Eq.~\ref{Eq:G1G2SpectralDecomposition} in terms of a sum of two spectral functions $\widetilde{G}_{1}(k_{x})\delta(k_{x}\!-\!k_{x}')$ and $\widetilde{G}_{2}(k_{x})\delta(k_{x}\!+\!k_{x}')$. Therefore the initial coherence function $\widetilde{G}(k_{x},\omega;k_{x}',\omega')$ now takes the form $\widetilde{G}(k_{x},\omega)\delta(\omega-\omega')\delta(|k_{x}|-|k_{x}'|)$, with $\widetilde{G}(k_{x},\omega)\!\rightarrow\!\widetilde{G}(k_{x})\delta(\omega-f(|k_{x}|))$. The $z$-dependence of the coherence function drops after introducing the spatio-temporal correlations. Therefore, the dimensionality of $\widetilde{G}$ has been reduced from 4D to 1D, and the propagation-invariant coherence function becomes
\begin{equation}\label{Eq:BroadbandSTCoherenceFunction}
G(x_{1},z,t_{1};x_{2},z,t_{2})\!=\!G_{1}(x_{1}\!-\!x_{2},t_{1}\!-\!t_{2})\!+\!G_{2}(x_{1}\!+\!x_{2},t_{1}\!-\!t_{2}),
\end{equation}
where $G_{j}(x,t)\!=\!\int\!dk_{x}\;\widetilde{G}_{j}(k_{x})e^{i(k_{x}x-\omega t)}$; $j\!=\!1,2$. The propagation-invariant intensity $I(x)\!=\!G(x,z,t;x,z,t)$ is given by
\begin{equation}\label{Eq:BroadbandIntensity}
I(x)=G_{1}(0,0)+G_{2}(2x,0),
\end{equation}
where the $G_{1}$-term contributes a constant background and the $G_{2}$-term contributes a spatial beam structure, in correspondence with Eq.~\ref{Eq:IntensityQuasiMonochromatic} for the quasi-monochromatic case. 

\begin{figure*}[t!]
\centering
\includegraphics[width=17.2cm]{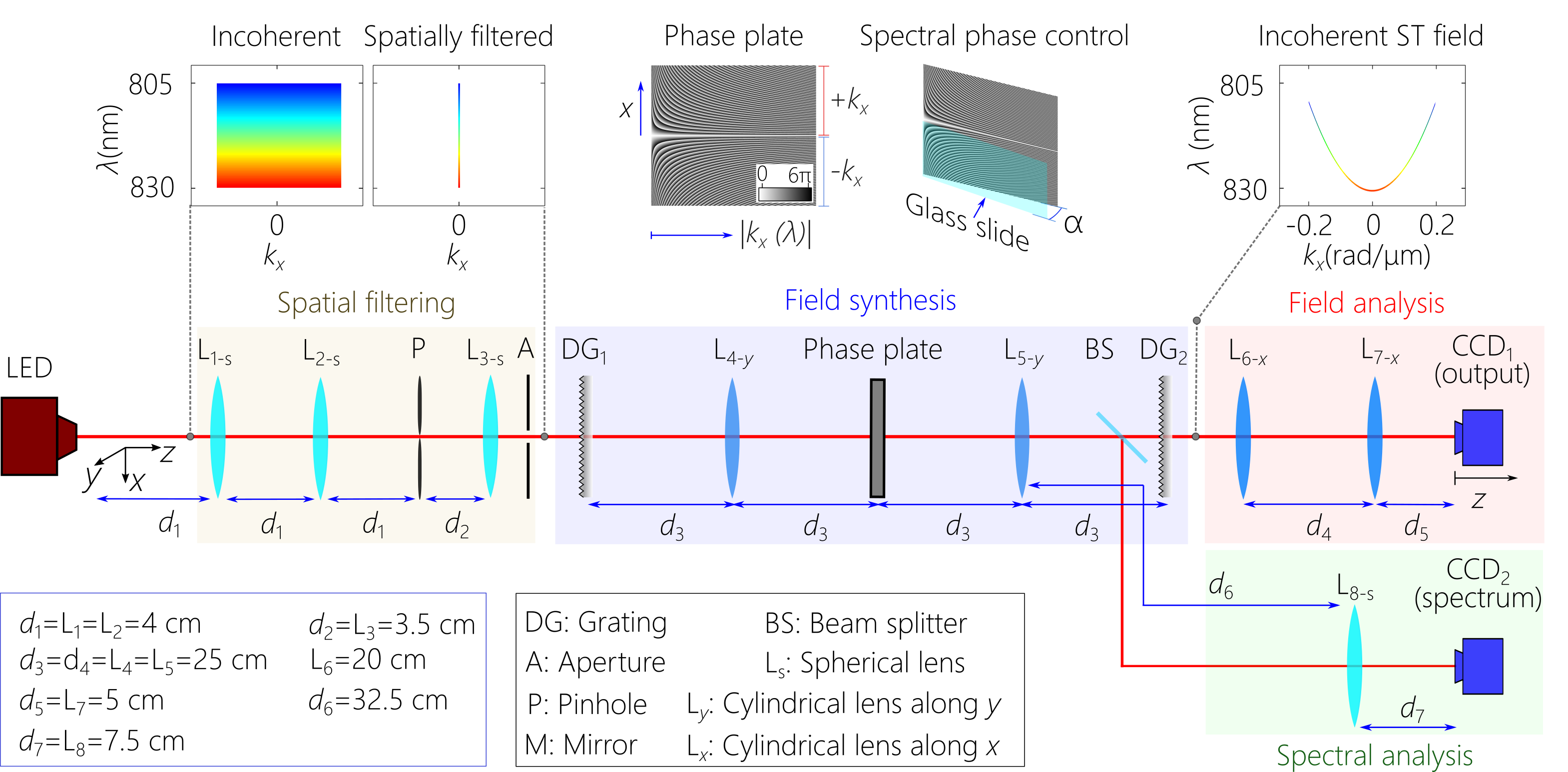}
\caption{\small{Optical setup for synthesizing incoherent ST fields. The arrangement consists of 4 sections delimited by the different-colored backgrounds: (i) spatial filtering via a $2f$-system comprising spherical lenses (L$_{2-\mathrm{s}}$ and L$_{3-\mathrm{s}}$) with a 200-$\mu$m-diameter pinhole P placed in the Fourier plane; (ii) field synthesis consisting of diffraction gratings DG$_1$ and DG$_2$, a transmissive phase plate, and cylindrical lenses (L$_{4-y}$ and L$_{5-y}$) for collimation; (iii) field analysis in physical space via a two-lens telescope (L$_{6-x}$ and L$_{7-x}$) and an axially scanning camera CCD$_1$; (iv) spatio-temporal spectral analysis in the $(k_{x},\lambda)$ domain via a $2f$-system along the $x$-axis (L$_{8-\mathrm{s}}$) to CCD$_2$. The gratings are used in reflection mode, but are depicted here in transmission mode for simplicity.}}
\label{Fig:Setup}
\end{figure*}

\begin{figure*}[t!]
\centering
\includegraphics[width=17.1cm]{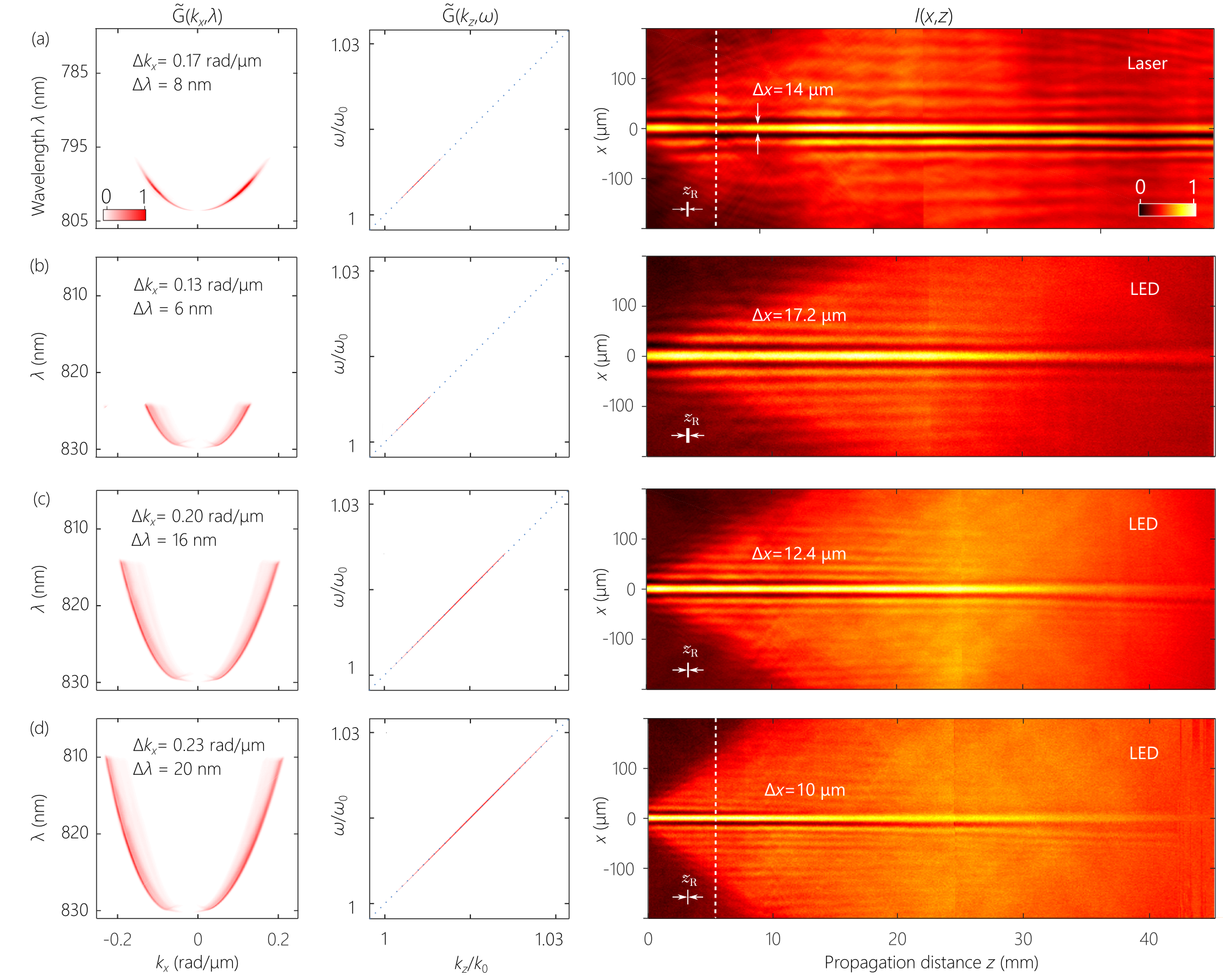}
\caption{\small{Measurements of the spatio-temporal spectrum and free-space propagation of broadband incoherent ST wave packets. The first column displays the spatio-temporal spectrum $\widetilde{G}(k_{x},\lambda)$, the second displays the same spectral density in the $(k_{z},\omega)$-plane, with $k_{z}$ normalized with respect to $k_{\mathrm{o}}\!=\!\tfrac{2\pi}{\lambda_{\mathrm{o}}}$ ($\lambda_{\mathrm{o}}\!=\!830$~nm, except in (a) where $\lambda_{\mathrm{o}}\!=\!803$~nm) and $\omega$ normalized with respect to $\omega_{\mathrm{o}}$, and the third column displays the time-averaged intensity of the field along the $z$-axis, $I(x,z)$ with $y\!=\!0$.  (a) Measurements carried out with a coherent mode-locked femtosecond Ti-Sapphire laser. (b-d) Measurements carried out with the incoherent field from a LED. From (b) to (d) we gradually increase the spatial bandwidth $\Delta k_{x}$, which results also in an increase in the temporal bandwidth $\Delta\lambda$. The spatial width of the field in the transverse plane $\Delta x$ accordingly decreases from (b) $\Delta x\!=\!17.2$~$\mu$m to (d) 10~$\mu$m. The dashed white lines in (a) and (d) identify the plane $z\!=\!z_{0}$ of the measurements in Fig.~\ref{Fig:Bright}, Fig.\ref{Fig:Dark}, and Fig.~\ref{Fig:Time}(b).}}
\label{Fig:ResultsBeamSize}
\end{figure*}

\section{Synthesis of incoherent space-time fields}

We now proceed to the experimental verification of the predictions presented above. The experimental setup employed in synthesizing incoherent ST fields is depicted schematically in Fig.~\ref{Fig:Setup}. The arrangement is divided into four sections: spatial filtering of the LED source; ST field synthesis; ST field spectral analysis; and ST field observation in physical space.

We make use of incoherent light from a broadband LED (Thorlabs, M810L3) that has a spectral bandwidth of $\approx\!29$~nm centered at a wavelength of 810~nm. The spatio-temporal spectrum corresponds to Fig.~\ref{Fig:Concept}(a); that is, a broadband incoherent fields with spatial and temporal bandwidths $\Delta k_{x}$ and $\Delta\lambda$, respectively. The beam is collimated with spherical lens L$_{1-\mathrm{s}}$ and spatially filtered with the combination of lenses L$_{2-\mathrm{s}}$ and L$_{3-\mathrm{s}}$ and a pinhole (diameter 300~$\mu$m), and subsequently filtered through aperture A ($25\!\times\!3$~mm$^{2}$). The spatio-temporal spectrum can now be approximated as a temporally incoherent plane wave, corresponding to Fig.~\ref{Fig:Concept}(b). To synthesize the ST field, we direct the field to a reflective diffraction grating DG$_{1}$ having a ruling with 1800~lines/mm and dimensions $25\!\times\!25$~mm$^2$ (Thorlabs GR25-1850) to spread the spectrum in space (incidence and reflection angles of $52^{\circ}$ and $42^{\circ}$, respectively). The first diffraction order is collimated with a cylindrical lens L$_{4-y}$ and directed to a transmissive phase plate of dimensions $25\!\times\!25$~mm$^2$. This transparent transmissive refractive phase plate was produced by gray-scale lithography (see Ref.~\cite{Wang15ProgPh}) and is designed to accommodate a temporal bandwidth extending to 40~nm. A 2D phase distribution [Fig.~\ref{Fig:Setup}, inset] is imparted to the impinging field, thus introducing correlations between $k_x$ and $\lambda$ corresponding to a tilt angle $\theta\!=\!45.3^{\circ}$, with the spread incoherent spectrum covering $\sim 18$~mm of its width ($\Delta k_{x}\!=\!0.24$~rad/$\mu$m) \cite{Kondakci18OE}.

\begin{figure}[t!]
\centering
\includegraphics[width=8.6cm]{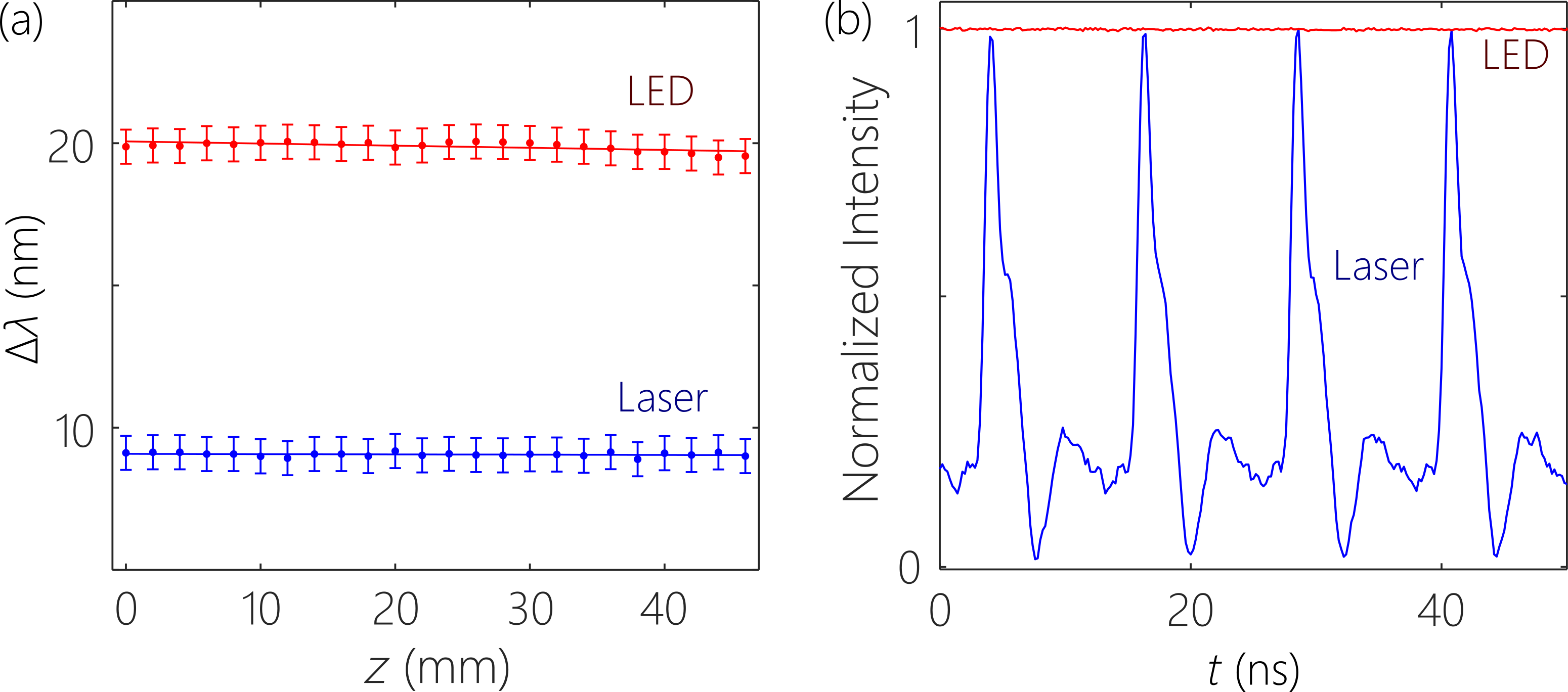}
\caption{\small{Measurement of the spectral bandwidth of ST field along propagation axis and intensity from fast detector at fixed axial plane. (a) FWHM of spectrum of ST field generated from coherent mode-locked femtosecond Ti-Sapphire laser (blue points with blue linear fit line) and broadband incoherent LED (red points with red linear fit line). (b) Normalized intensity of the coherent and incoherent ST field.}}
\label{Fig:Time}
\end{figure}

The field transmitted through the phase plate is directed to a second grating DG$_2$ (identical to DG$_{1}$) through a cylindrical lens L$_{5-y}$ (identical to L$_{4-y}$), whereupon the input spectrum is reconstituted and the incoherent ST field is formed. The ST field is subjected to two measurement modalities: characterization in physical space $(x,z)$ by a CCD$_{1}$ camera (The ImagingSource, DMK 27BUP031) moving along the $z$-direction to capture the time-averaged intensity profile $I(x,y,z)$; and characterization in the spatio-temporal spectral domain or $(k_{x},\lambda)$-plane recorded by CCD$_2$ (The ImagingSource, DMK 27BUJ003) after performing a Fourier transform along the $x$-axis (lens L$_{8-\mathrm{s}}$) and spreading the temporal spectrum along $y$ (combination of lenses L$_{5-y}$ and L$_{8-\mathrm{s}}$). The spectral uncertainty of ST field is limited by the spectral resolution of the grating and phase plate and by the spatial filtering of the LED light.

This experimental arrangement was developed in our previous work on coherent pulsed ST wave packets. This approach has enabled us to overcome the decades-old challenge of verifying that the group velocity of ST wave packets can be arbitrary in free space \cite{Kondakci18unpub}. Previous experimental results did not show deviations from $c$ by more than $0.1\%$, whereas our approach has enabled the tuning of the group velocity of ST wave packets in free space from $-4c$ to $30c$ \cite{Kondakci18unpub}. Here, we have extended the utility of this novel approach to incoherent light, thus demonstrating that ultrafast-pulse modulation techniques can be exploited in manipulating the spatio-temporal coherence of an optical field. Consequently, we are now able to experimentally synthesize never-before-observed hypothesized incoherent non-diffracting field configurations at unprecedented precision.

\section{Characterization of incoherent space-time fields}

\subsection{Measurements of the spatio-temporal spectra and evolution of the intensity in physical space}

\begin{figure*}[t!]
\centering
\includegraphics[width=17.2cm]{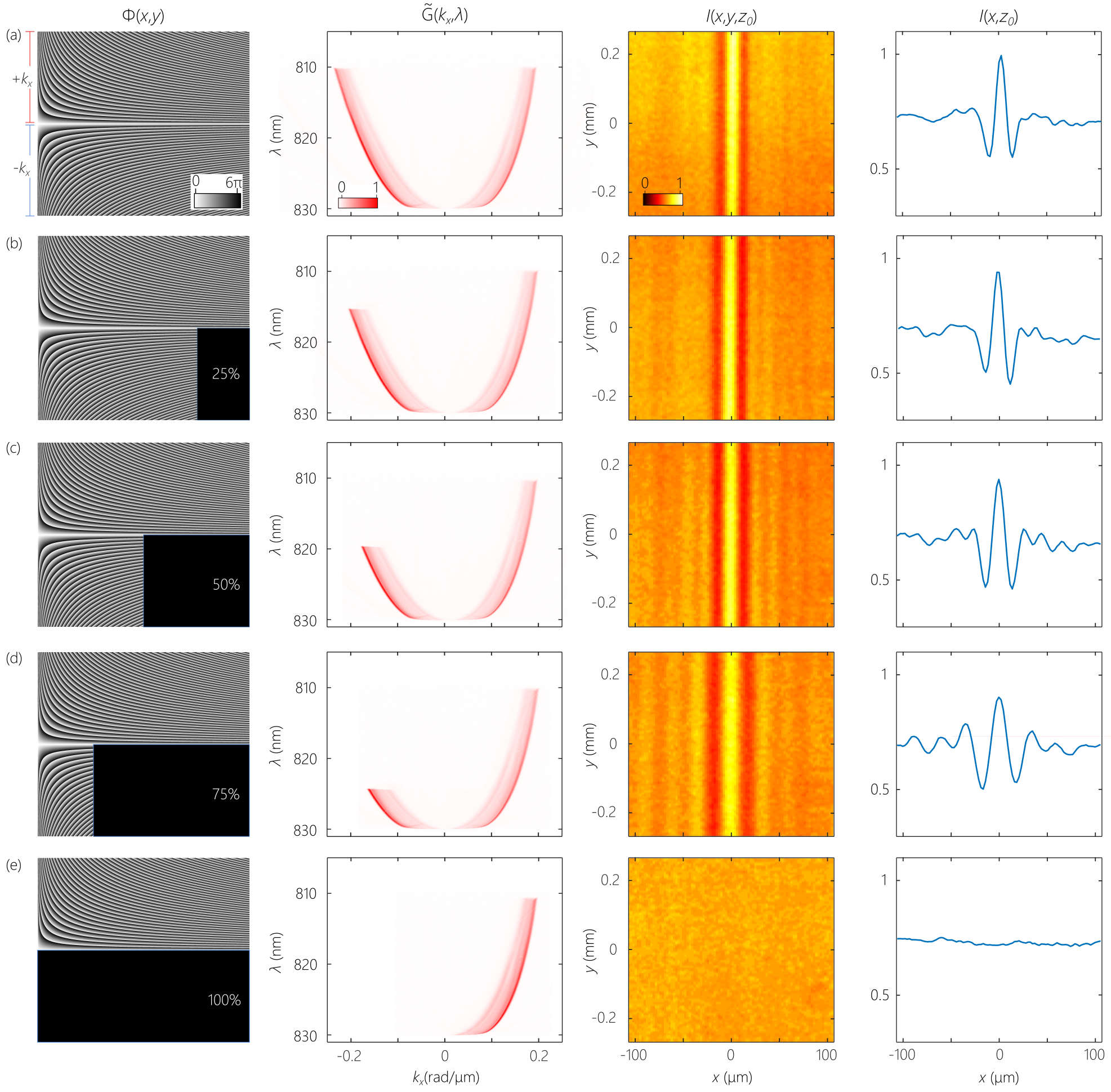}
\caption{\small{Controlling the ratio of the constant background $G_{1}(0)$ and the diffraction-free narrow spatial feature $G_{2}(2x)$ through spatio-temporal spectral filtering. The first column shows the portion of the phase distribution $\Phi$ on the phase plate that is unblocked; the second column shows the spatio-temporal spectrum $\widetilde{G}(k_{x},\omega)$, the third column the time-averaged transverse intensity $I(x,y,z_{0})$ captured by CCD$_{1}$ at the plane $z\!=\!z_{0}$, which is identified by the dashed white line in the plot of $I(x,z)$ in Fig.~\ref{Fig:ResultsBeamSize}(d); and the last column shows the 1D propagation-invariant intensity distribution $I(x,z_{0})$ resulting from integration over $y$.}}
\label{Fig:Bright}
\end{figure*}

We first confirm that the phase plate has correctly introduced the desired spatio-temporal spectral correlations corresponding to a spectral tilt angle of $\theta\!=\!45.3^{\circ}$, which confines the locus of the spatio-temporal spectrum to a hyperbola on the light-cone; Fig.~\ref{Fig:ResultsBeamSize}. We first carry out an experiment with a coherent mode-locked femtosecond Ti:Sapphire laser (Tsunami, Spectra Physics), which has a central wavelength of $\sim800$~nm and bandwidth of $\sim8$~nm, corresponding to a pulse width of $\sim100$~fs. After spreading the laser spectrum via grating DG$_1$, the bandwidth covers only a portion of the phase plate, resulting in a reduced spatial spectrum, which is in excellent agreement with the targeted correlation function between $k_{x}$ and $\lambda$ [Fig.~\ref{Fig:ResultsBeamSize}(a)]. We also plot the spatio-temporal spectrum in the $(k_{z},\tfrac{\omega}{c})$ through the appropriate transformation from the $(k_{x},\lambda)$-plane and normalizing $k_{z}$ and $\omega$ with respect to $k_{\mathrm{o}}$ and $\omega_{\mathrm{o}}$, respectively, where $k_{\mathrm{o}}\!=\!\tfrac{2\pi}{\lambda_{\mathrm{o}}}$ and $\lambda_{\mathrm{o}}\!=\!800$~nm. The measured curve confirms to the expected linear relationship. The resulting ST wave packet has a transverse spatial width of $\Delta x\!\approx\!14$~$\mu$m, and is confirmed to be quasi-nondiffracting over a distance of at least $\sim30$~mm. Note that the Rayleigh range $z_{\mathrm{R}}$ of a beam of width 14~$\mu$m at a wavelength of 800~nm is only $\sim0.192$~mm. The propagation distance observed here is thus $\sim150z_{\mathrm{R}}$. 

We next proceed to use the broadband LED in lieu of the pulsed laser. The available temporal bandwidth is $\Delta\lambda\!\approx\!29$~nm, but after spreading the spectrum $\Delta\lambda\!\approx20$~nm is intercepted by the phase plate. By blocking a portion of the phase plate, additional temporal spectral filtering can be applied. We start by uncovering approximately the same bandwidth as the Ti:Sapphire laser ($\Delta\lambda\!\approx\!8$~nm); Fig.~\ref{Fig:ResultsBeamSize}(b). The measured spatio-temporal spectrum in the $(k_{x},\lambda)$ and $(k_{z},\tfrac{\omega}{c})$ plots are similar to those of the Ti:Sa laser in Fig.~\ref{Fig:ResultsBeamSize}(a), except for a shift along the $\lambda$-axis, a larger spectral uncertainty $\delta\lambda$ caused by the limited spatial filtering of the LED light, and the fact that all the spectral amplitudes are now mutually incoherent. The results of observing the broadband incoherent ST field with CCD$_{1}$ reveal a narrow spatial feature of transverse width $\sim17$~$\mu$m atop a broad background propagating for an axial distance of $\sim\!25$~mm (corresponding to $\sim90z_{\mathrm{R}}$).

We further increase the width of the spatial and temporal bandwidths of incoherent ST field by uncovering a wider extent of the phase plate. In Fig.~\ref{Fig:ResultsBeamSize}(c) we have $\Delta\lambda\!=\!16$~nm, $\Delta k_{x}\!=\!0.2$~rad/$\mu$m, $\Delta x\!=\!12.4$~$\mu$m, and a propagation length corresponding to $\sim200z_{\mathrm{R}}$; and in Fig.~\ref{Fig:ResultsBeamSize}(d) we have $\Delta\lambda\!=\!20$~nm, $\Delta k_{x}\!=\!0.23$~rad/$\mu$m, and $\Delta x\!=\!10$~$\mu$m, propagating $\sim300z_{\mathrm{R}}$. In all cases, the time averaged intensity $I(x,z)$ at any plane $z$ takes the form of a sum of a constant background and a narrow spatial feature, corresponding to the terms $G_{1}(0)$ and $G_{2}(2x)$, respectively in Eq.~\ref{Eq:BroadbandIntensity}, with the width of the narrow feature $G_{2}(2x)$ inversely proportional to the spatial bandwidth $\Delta k_{x}$ of $\widetilde{G}_{2}(k_{x})$.

Because we have characterized the broadband incoherent ST field before its reconstitution after DG$_2$, we carry out a spectral measurement downstream after the formation of the field. We measure the temporal bandwidth $\Delta\lambda$ of the ST field along the propagation axis $z$. Instead of monitoring the spatial intensity distribution by scanning CCD$_1$ along $z$ (Fig.~\ref{Fig:ResultsBeamSize}, third column), we scan a spectrometer (Thorlabs CCS175) to determine the $z$-dependence of the temporal bandwidth. The experimental results shown in Fig.~\ref{Fig:Time}(a) correspond to the coherent ST wave packet in Fig.~\ref{Fig:ResultsBeamSize}(a) and and the incoherent ST wave packet in Fig.~\ref{Fig:ResultsBeamSize}(d). In both cases, the temporal bandwidth is stable with propagation in agreement with the invariance of the intensity distribution. 

Despite the similarity so far between coherent ST wave packets and incoherent ST fields, a fundamental distinction exists: the coherent ST wave packet is fixed, while incoherence renders ST wave packets quasi-CW. This is confirmed in Fig.~\ref{Fig:Time}(b) where a fast photodiode detector (Thorlabs FDS010; $\sim\!1$~ns response time) connected to 500~MHz digital oscilloscope (Tektronix TDS3054B) captures the field at the axial plane $z_{0}\!=\!5$~mm, which is identified in Fig.~\ref{Fig:ResultsBeamSize}(a) and Fig.~\ref{Fig:ResultsBeamSize}(d) by a dashed white line. 

\subsection{Modulating the correlation amplitude of symmetric spatial frequency pairs}

The amplitude of the narrow spatial feature $G_{2}(2x)$ depends on the magnitude of the correlation of the spatial frequency pairs $\pm k_{x}$, and at best is equal to the background $G_{1}(0)$. The amplitude of $G_{2}(2x)$ can thus be controlled with respect to the background by reducing the magnitude of $\widetilde{G}_{2}(k_{x})$, which we achieve by gradually blocking the negative spatial frequencies, as shown in the first column of Fig.~\ref{Fig:Bright}. Note that the top half of the phase plate encodes the positive spatial frequencies, while the lower half encodes the negative ones. By placing a physical beam block in front of the \textit{lower} half of the phase plate, we can adjust $\widetilde{G}_{2}(k_{x})$. In Fig.~\ref{Fig:Bright} we show the results of gradually blocking the negative spatial frequencies (resulting in an asymmetric spatio-temporal spectrum) until only positive spatial frequencies enter into the formation of the incoherent ST field. Note that the temporal bandwidth $\Delta\lambda$ remains constant throughout. 

The measurements given in Fig.~\ref{Fig:Bright} correspond to the configuration in Fig.~\ref{Fig:ResultsBeamSize}(d) where maximal temporal and spatial bandwidths are utilized, resulting in a spatial width of $\Delta x\!\sim\!10$~$\mu$m. Observations of the intensity are made at the axial plane $z_{0}\!=\!5$~mm (identified in Fig.~\ref{Fig:ResultsBeamSize}(d) by a dashed white line). As the negative half of the spatial spectrum is gradually blocked, we observe a drop in the amplitude of the narrow spatial feature $G_{2}(x,0)$ with respect to the background term $G_{1}(0,0)$, until it completely vanishes when the negative half of the spectrum is eliminated altogether, leaving behind only an incoherent constant-intensity background, as shown in Fig.~\ref{Fig:Bright}(a) through ~\ref{Fig:Bright}(e). Note that $\Delta x$ increases as expected through the spatial filtering process.

\subsection{Modulating the correlation phase of symmetric spatial frequency pairs: `dark' and `bright' fields}

\begin{figure}[t!]
\centering
\includegraphics[width=8.6cm]{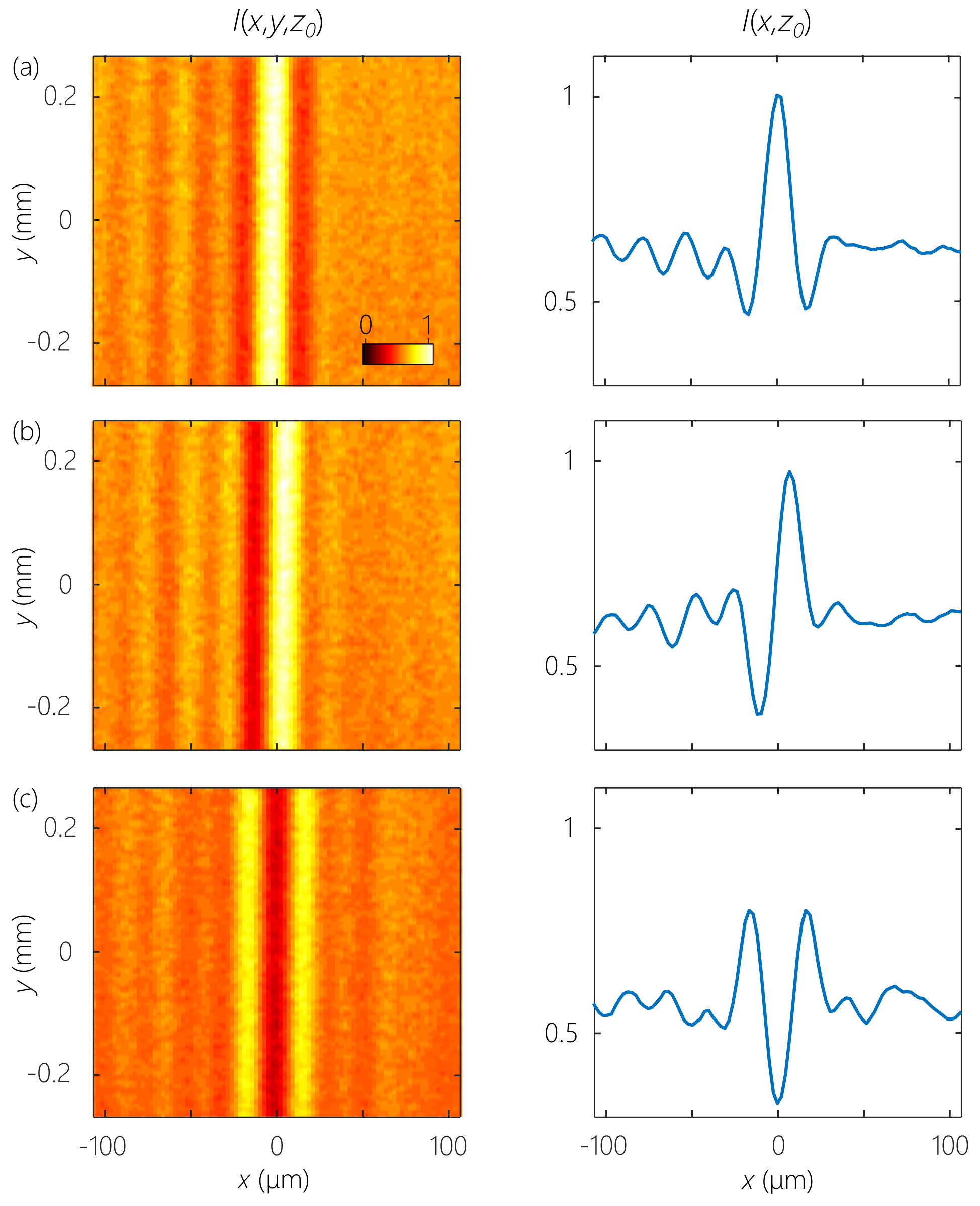}
\caption{\small{`Bright' and `dark' diffraction-free incoherent ST fields. The first column shows the transverse time-averaged intensity $I(x,y,z_{0})$, and the second column shows a 1D intensity $I(x,z_{0})$ resulting from integration over $y$. (a) A \textit{bright} incoherent ST field is formed when the spatial spectral function $\widetilde{G}_{2}(k_{x})$ is even. (b) An intermediate bright-dark incoherent ST field in which the relative phase between the positive and negative halves of the spatial spectrum is $\sim\tfrac{\pi}{2}$. (c) A \textit{dark} incoherent ST field is formed when the spatial spectral function $\widetilde{G}_{2}(k_{x})$ is odd, corresponding to a $\pi$-step in phase between the two halves of the spectrum.}}
\label{Fig:Dark}
\end{figure}

We next introduce a phase difference between the positive and negative halves of the spatial spectrum, thus modulating the phase of the $\widetilde{G}_{2}(k_{x})$ term. To introduce such a phase difference, we make use of a glass microscope slide (thickness $\sim100$~$\mu$m; Fisher Scientific 12-542C, area $25\times25$~mm$^{2}$), which we place in front of the  phase plate covering the lower half corresponding to the negative spatial frequencies and observe the time-averaged intensity at the axial plane $z_{0}\!=\!5$~mm. If the microscope slide is parallel to the phase plate, then a constant phase is applied to the negative half of the spectrum. By tilting the slide, this relative phase can be varied; Fig.~\ref{Fig:Dark}. The zero-phase condition is identified by observing the maximal amplitude for the narrow spatial feature $G_{2}(2x,0)$; Fig.~\ref{Fig:Dark}(a). On the other hand, introducing a $\pi$-step in phase along the $k_{x}$-axis is associated with a minimum value at the field center below the constant background; Fig.~\ref{Fig:Dark}(c). The intermediate configuration of $\approx\tfrac{\pi}{2}$-step in phase along the $k_{x}$-axis results in a spatial feature combining a peak and a dip \ref{Fig:Dark}(b). Note that $\Delta x$ of the spatial feature does not change here, in contrast to the case of \textit{filtering} the spatial spectrum presented above.

\section{Discussion}

In 1979, Berry proved that there are no 1D monochromatic diffraction-free beams except for those whose profile conforms to an Airy function \cite{Berry79AMP}, in which case the beam trajectory traces a parabolic curve away from the optical axis \cite{Siviloglou07OL}. Lifting the monochromaticity constraint, 1D propagation-invariant coherent wave packets in the form of light sheets with arbitrary spatial profiles and finite temporal bandwidth can be synthesized \cite{Kondakci17NP} -- including non-accelerating Airy beams that travel in straight lines \cite{Kondakci18PRL}. We have extended this concept here to \textit{incoherent} fields by studying the impact of introducing tight spatio-temporal \textit{spectral} correlations into broadband incoherent light. This work paves the way to the investigation of incoherent ST fields as an example of light structured in space and time via a spatio-temporal Fourier-optics strategy. Indeed, incoherent ST fields may be considered the polar \textit{opposite} of cross-spectrally pure light \cite{Mandel61JOSA,Mandel76JOSA}: instead of spatio-temporal \textit{separability} we have tight correlations that result in \textit{non}-separability. In this paper, we have confirmed that the tight spatio-temporal spectral correlations introduced into the field result in a diffraction-free broadband field confirmed through intensity measurements in physical space. Future work will characterize the spatio-temporal coherence function $G(x_{1},z,t_{2};x_{2},z,t_{2})$.

We have studied incoherent ST fields in the limit of delta-function correlations between spatial and temporal frequencies, whereupon $\omega\!=\!f(|k_{x}|)$, and the functional form of $f(|k_{x}|)$ is a conic section determined by the title angle $\theta$ of the hyperplane intersecting a light-cone \cite{Yessenov18PRA}. In this idealized scenario, the diffraction-free propagation length is formally infinite: the evolution of the coherence function and hence also the intensity profile are independent of $z$. Introducing a finite spectral uncertainty $\delta\omega$ [Fig.~\ref{Fig:Concept}(d)] results in \textit{partially coherent} ST fields. This can be incorporated into the theory by replacing the delta-function correlation with the more-realistic assumption of a narrow function correlation $g(\omega-f(|k_{x}|))$, where $g$ has a width of $\delta\omega$ and is centered for each value of $\omega$ at the idealized value $f(|k_{x}|)$. This seems to be a viable approach to studying partially coherent ST fields. The impact of increasing $\delta\omega$ is to transition to a partially coherent field \textit{and} to render the diffraction-free length finite. Accompanying these changes is a reduction of the amplitude of the background term \cite{Kondakci16OE}. At the limit of large $\delta\lambda$, the incoherent field becomes separable in space and time, resulting in cross-spectrally pure light. The question remains whether it is possible to synthesize a diffraction-free partially coherent field. 

Another important question is in regards to the limits of the synthesis process described here. In principle, one can exploit very large temporal bandwidths $\Delta\lambda$ (white light) and also large spatial bandwidths $\Delta k_{x}$ to produce even narrower spatial features. This will likely require considering an electromagnetic field in lieu of the scalar treatment adopted here.

Several avenues will be pursued in future work. One may now investigate effects associated previously with coherent ST wave packets in the context of incoherent light, such as self-healing \cite{Kondakci18OL}, the propagation for extended distances \cite{Bhaduri18OE}, among other possibilities. Second, we will study the synthesis of incoherent ST fields having structured spatial profiles in both transverse dimensions (2D instead of 1D as demonstrated here).In this paper, we have examined the intensity of the ST field to confirm its diffraction-free behavior. Future experiments will focus on measuring the full coherence function $G$. Future work will concentrate on performing coherence measurements on these ST fields. A double-slit interference experiment on a broadband incoherent ST field described by Eq.~\ref{Eq:BroadbandSTCoherenceFunction} at symmetric points $\pm x$ about the origin gives an interference pattern whose visibility is proportional to the degree of spatial coherence,
\begin{equation}
G(x,t;-x,t+\tau)=G_{1}(2x,\tau)+G_{2}(0,\tau).
\end{equation}
Spatial and temporal incoherence in this context retain the usual meanings of the coherence function dropping to zero after some spatial or temporal separation, respectively. There is a complementarity between the roles played by $G_{1}$ and $G_{2}$ in the intensity and coherence domains. When measuring the \textit{intensity}, $G_{1}$ contributes a constant background, while $G_{2}$ contributes a narrow spatial feature. Alternatively, when measuring the \textit{coherence function}, $G_{1}$ provides a finite-width coherence feature, while $G_{2}$ provides a constant pedestal, which can be reduced or even eliminated with the approach used in Fig.~\ref{Fig:Bright}. Uniquely, spatial and temporal coherence are not independent. Because the spatial and temporal frequencies are tightly correlated, the spatial and temporal bandwidths, $\Delta k_{x}$ and $\Delta\omega$, respectively, are also related via $\tfrac{\Delta\omega}{\omega_{\mathrm{o}}}\!=\!|f(\theta)|\tfrac{(\Delta k_{x})^{2}}{2k_{\mathrm{o}}^{2}}$, where $f(\theta)\!=\!\tfrac{1}{\cot{\theta}-1}$ and $\theta$ is the spectral tilt angle (Eq.~\ref{Eq:CorrelationFunction}) \cite{Yessenov19OL}. The temporal coherence length and transverse spatial coherence width (inversely related to $\Delta\omega$ and $\Delta k_{x}$) are thus tightly related, and there proportionality is determined by $\theta$.

Finally, although it might be expected that the time-averaged intensity measurements must always resemble the time-averaged intensity of a coherent pulsed ST wave packet having the same spectrum, this is not always the case. In our experimental synthesis process, the field amplitudes associated with the spatial-frequency pair $\pm k_{x}$ assigned to the same frequency $\omega$ are mutually coherent. By introducing, for example, a random phase between the field amplitudes for $\pm k_{x}$, the $G_{2}$-term representing the central spatial feature vanishes. In that case, the time-averaged intensity differs between the coherent and incoherent counterparts.

In the context of coherence measurements, two intriguing questions can be addressed. First, eliminating $G_{2}$ (by blocking the negative half of the spatial spectrum) produces a tilted coherence front -- the incoherent counterpart to tilted pulse fronts \cite{Torres10AOP} in which the pulse intensity front is tilted with respect to the phase front and thus also to the direction of propagation \cite{Wong17ACSP2,Kondakci18ACSP}. Can the phenomenon of tilted \textit{coherence} fronts be observed? Second, what is the physical significance of the tilt angle $\theta$ and hence the group velocity $v_{\mathrm{g}}\!=\!c\tan{\theta}$ in the context of incoherent light? Equation~\ref{Eq:PropagationOfBroadbandCoherence} indicates that the delay $t_{1}-t_{2}$ required to compensate for measuring coherence in two different planes $z_{1}$ and $z_{2}$ is not related to the separation $z_{1}-z_{2}$ through the speed of light in vacuum. Instead, the delay corresponds to propagation at a group velocity $v_{\mathrm{g}}$. Varying the tilt angle $\theta$ by modifying the phase pattern imparted to the spread spectrum in our experiment is thus expected to lead to subluminal and superluminal -- or even negative-velocity -- propagation of coherence in free space, none of which have been observed to date. Finally, the impact of the spectral uncertainty on the axial propagation of the coherence function needs to be assessed in light of our recent demonstrations that it is the spectral uncertainty that determines the diffraction-free length of coherent ST wave packets \cite{Kondakci19OL,Yessenov19OL,Bhaduri19OL}.

\section{Conclusions}

In conclusion, we have presented the first experimental demonstration of an incoherent broadband field in which the spatial and temporal degrees of freedom are structured together in a precise and controllable manner resulting in non-diffracting propagation. This is made possible by exploiting phase-only synthesis techniques that have previously been confined to ultrafast optics, in contrast to previous results that utilized coherent light (monochromatic or pulsed) or monochromatic lasers in which spatial incoherence was introduced via a diffuser -- all of which relied on traditional approaches for preparing Bessel beams. Experiments exploiting a broadband incoherent field reached the opposite conclusion to ours: that spatial and temporal incoherence inevitably lead to a diminishing of the diffraction-free propagation distance. We have shown here that the same diffraction-free distance associated with a coherent femtosecond laser is observed with a broadband LED as long as precise spatio-temporal correlations are introduced into the spectrum of the incoherent field. By shaping the 20-nm spectral bandwidth of the LED at an unprecedented precision of $\sim25$~pm, we have observed a diffraction-free length of more than 30~mm for a bright or dark spatial feature of width 10~$\mu$m in an incoherent field, in excess of $\sim300\times$ the Rayleigh range. The diffraction-free incoherent fields described here provide opportunities for new applications in optical imaging, light-sheet microscopy, and spectroscopy. Our work thus demonstrates that the methodologies developed for ultrafast optics can be exploited in the realm of incoherent optics for the synthesis of new field configurations and can enable the full investigation of partial coherence in non-diffracting fields. We expect this aspect of our work to have an impact on optical coherence far beyond the specific experiment we report on here.

%In conclusion, we have experimentally demonstrated that broadband incoherent diffraction-free ST fields can be synthesized \textcolor{blue}{starting from LED light} using a novel optical configuration recently introduced in the context of propagation-invariant coherent pulsed beams. Using this strategy, diffraction-free incoherent ST fields in the form of light sheets are produced by introducing the appropriate spatio-temporal spectral correlations. To the best of our knowledge, this is the first realization of a broadband incoherent field in which precise and versatile control is exercised over its spatio-temporal spectral content through the introduction of tight correlations between the spatial and temporal degrees of freedom.
%\textcolor{blue}{By introducing techniques from ultrafast optics combined with Fourier optics, we shape the 20-nm-bandwidth LED light at an unprecedented precision of $\sim\!25$~pm, leading to propagation distances $\sim\!300\!\times$ the Rayleigh of a beam having the same initial transverse width and a non-diffraction null in the incoherent ST field. The appropriation of such techniques into the context of optical coherence can have an impact that extends beyond the particular experiment presented here.} The diffraction-free incoherent fields described here provide opportunities for new applications in optical imaging, light-sheet microscopy, and spectroscopy.

%\vspace{4pt}
%\textbf{Acknowledgments}. XXXXX

\vspace{4pt}
\textbf{Funding}. U.S. Office of Naval Research (ONR) contract N00014-17-1-2458. U.S. Office of Naval Research (ONR) contract N66001-10-1-4065.

%\bibliography{Diffraction}

%\bibliographyfullrefs{Diffraction}

\end{document}